\documentclass[a4paper,twocolumn,aps,10pt,english,amsmath,amssymb,showpacs,floatfix,notitlepage,nofootinbib,superscriptaddress,longbibliography]{revtex4-1}
\usepackage{graphicx}
\usepackage{mathrsfs}
\usepackage[utf8]{inputenc}
\usepackage[unicode=true,pdfusetitle,
	bookmarks=true,bookmarksnumbered=false,bookmarksopen=false,
breaklinks=false,pdfborder={0 0 1},colorlinks=false]{hyperref}
\hypersetup{colorlinks,citecolor=blue,linkcolor=blue}
\begin{document}

\title{High order flow harmonics of identified hadrons in 2.76 A TeV Pb+Pb collisions}

\author{Hao-jie Xu}
\affiliation{Department of Physics and State Key Laboratory of Nuclear Physics and
Technology, Peking University, Beijing 100871, China}
\author{Zhuopeng Li}
\affiliation{Department of Physics and State Key Laboratory of Nuclear Physics and
Technology, Peking University, Beijing 100871, China}
\author{Huichao Song}
\email{Huichao Song: huichaosong@pku.edu.cn}
\affiliation{Department of Physics and State Key Laboratory of Nuclear Physics and
Technology, Peking University, Beijing 100871, China}
\affiliation{Collaborative Innovation Center of Quantum Matter, Beijing 100871, China}
\affiliation{Center for High Energy Physics, Peking University, Beijing 100871, China}

\date{\today}

\begin{abstract}
Using {\tt iEBE-VISHNU} hybrid model with the {\tt AMPT} initial conditions,
we study the higher order flow harmonics of identified hadrons in
2.76 A TeV Pb+Pb collisions. Comparison with the recent ALICE measurements at 20-30\% centrality shows that 
our calculations nicely describe the data below 2 GeV, especially for the $v_2$, $v_3$ 
and $v_4$ mass-orderings among pions, kaons and protons. We  also extended the calculations to other centrality bins, 
which presents similar mass-ordering patterns for these flow harmonics as the ones observed at 20-30\% centrality. 
In the later part of this article, we explore the development of $v_n$ mass ordering/splitting during the hadronic evolution through
the comparison runs from {\tt iEBE-VISHNU} hybrid model and pure hydrodynamics
with different decoupling temperatures.
\end{abstract}


\maketitle

\section{Introduction}
\label{introduction}

It is widely believed that the quark-gluon plasma (QGP)
has been created in relativistic heavy ion collisions at Relativistic
Heavy Ion Collider (RHIC) and the Large Hadron Collider (LHC)~\cite{Rev-Arsene:2004fa,Gyulassy:2004vg,Muller:2006ee,reviews}.
The observation of strong collective flow and the successful description from hydrodynamics
reveal that the QGP is strongly coupled and behaves as an almost perfect
liquid~\cite{Gyulassy:2004vg,Muller:2006ee,reviews}.
With the development of viscous hydrodynamics and the related hybrid models, the elliptic flow, triangular flow
and other higher order flow harmonics have been widely used to study and extract the transport properties of
the QGP~\cite{Romatschke:2009im,Heinz:2013th,Gale:2013da,Song:2012ua,Romatschke:2007mq,Song:2007fn,
Dusling:2007gi,Song:2010mg,Niemi:2011ix,Song:2009rh,Bozek:2009dw,Dusling:2011fd,Noronha-Hostler:2013gga,Ryu:2015vwa,Karpenko:2015xea,Schenke:2010rr,Gale:2012rq,Niemi:2015qia}.
These higher order flow harmonics also reveal that the created QGP fireball
strongly fluctuates event by event, which finally transforms into final state
correlations of the produced hadrons after the collective
expansion~\cite{Schenke:2010rr,Gale:2012rq,Niemi:2015qia,
Petersen:2010cw,Qin:2010pf,Holopainen:2010gz,Qiu:2011iv,Luzum:2013yya,ALICE:2011ab,ATLAS:2012at,Adare:2011tg,Adamczyk:2013waa}.
Meanwhile, other flow measurements, such as  $v_n$ in ultra-central collisions~\cite{Luzum:2012wu}, the distributions of event-by-event flow harmonics~\cite{Aad:2013xma}, the correlations between flow angles~\cite{Aad:2014fla}, and the correlations between flow harmonics of different order~\cite{Aad:2015lwa,Zhou:2015slf} etc., provide more information on the initial state fluctuations, which help to constrain the initial conditions of hydrodynamic simulation for a precise extraction of the QGP transport properties.

Besides these flow data of all charged hadrons, the elliptic flow of identified hadrons have also
been extensively measured and studied at RHIC and the
LHC~\cite{Adams:2003am,Adler:2003kt,Adams:2005zg,Abelev:2007qg,Abelev:2014pua,Song:2010mg,Song:2013qma,Zhu:2015dfa,
Zhu:2015eea}. As a typical feather of the collective expansion, the mass ordering of elliptic flow among various hadron species
has been observed in different colliding systems~\cite{Adams:2003am,Adler:2003kt,Adams:2005zg,Abelev:2007qg,Abelev:2014pua},
which reflects the information of the hadronic evolution after
the phase transition. Recently, the ALICE collaboration has measured the
higher order flow harmonics of identified hadrons in ultra-central and semi-central Pb+Pb collisions
at $\sqrt{s_{NN}}$ = 2.76 TeV~\cite{ALICE-QM2015}, which showed that $v_3(p_T)$ and $v_4(p_T)$  present similar
mass ordering patterns among pions, kaons and protons as
observed in the elliptic flow. It is thus the right time to systematically
study these new flow measurements  with sophisticated model
calculations. In this paper, we will implement {\tt iEBE-VISHNU} hybrid model that couples 2+1-d viscous hydrodynamics to the
hadron cascade model, together with the fluctuating {\tt AMPT} initial conditions, to calculate
 $v_2(p_T)$, $v_3(p_T)$ and $v_4(p_T)$ of pions, kaons and protons in
2.76 A TeV Pb+Pb collisions. We will compare our model calculations with the recent ALICE measurements at 20-30\%
centrality and make predictions for other centrality bins. In the later part of this article, we will explore
the development of $v_n$ mass orderings/splittings during the hadronic evolution through
the comparison runs from {\tt iEBE-VISHNU} and from pure hydrodynamics decoupled at
different temperatures.

\begin{figure*}
	\begin{centering}
		\includegraphics[scale=0.9]{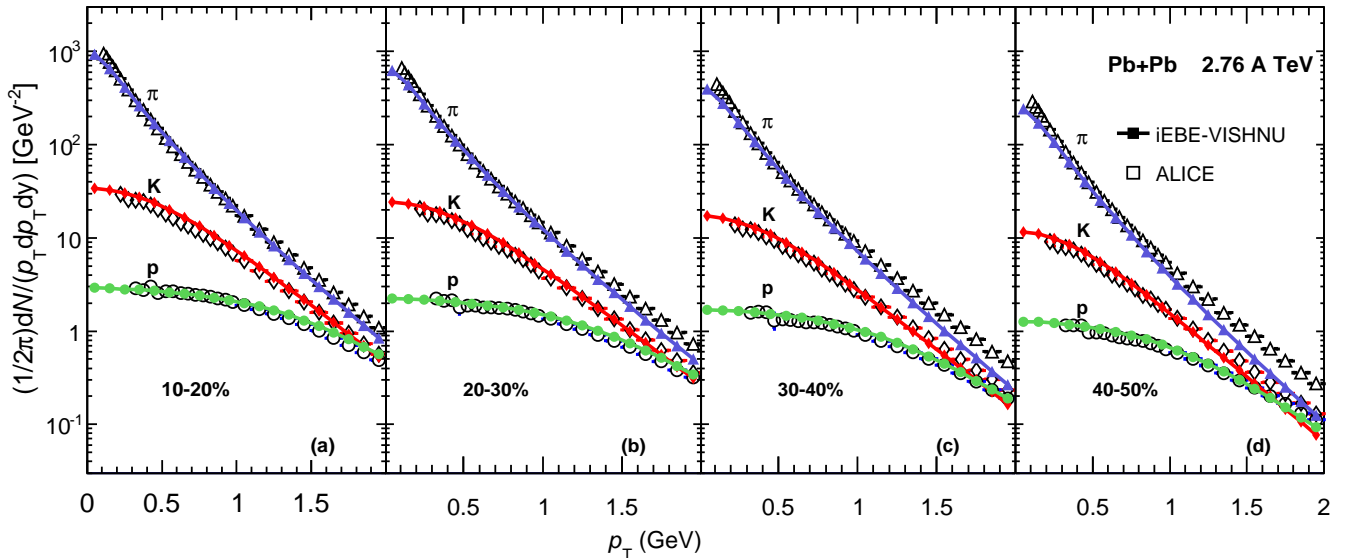}
	\end{centering}
    \vspace{-5mm}
	\caption{(Color online) $p_{T}$-spectra of pions, kaons and protons in 2.76 A TeV Pb+Pb collisions at $10-20\%$, $20-30\%$, $30-40\%$ and $40-50\%$ centralities. The experimental data are taken from the ALICE paper~\cite{Abelev:2013vea}, and the theoretical curves are calculated from {\tt{iEBE-VISHNU}} with {\tt{AMPT}} initial conditions.
	\label{fig:ptspectra} }
\end{figure*}

\section{Setup}
\label{setup}

{\tt{iEBE-VISHNU}}~\cite{Shen:2014vra} is an event-by-event version of {\tt VISHNU} hybrid model~\cite{Song:2010aq}
which combines (2+1)-dimensional viscous hydrodynamics ({\tt{VISH2+1}})~\cite{Song:2007fn,Song:2009gc}
to describe the QGP expansion with the hadron cascade model ({\tt{UrQMD}})~\cite{Bass:1998ca,Bleicher:1999xi}
to simulate the evolution of the hadronic matter. The transition between the macroscopic and microscopic approaches is
determined by a switching temperature $T_{sw}$, which is set to $165 \mathrm{MeV}$ as used in
Ref.~\cite{Song:2010mg,Song:2013qma,Zhu:2015dfa}. For the
hydrodynamic simulations, we input an equation of state (EoS) s95p-PCE constructed from combing the lattice EoS
for the QGP phase with a partially chemical equilibrium EoS for the hadron resonance gas phase~\cite{Huovinen:2009yb,Shen:2010uy}.

In the published version of {\tt{iEBE-VISHNU}}~\cite{Shen:2014vra}, the initial entropy density (energy density) profiles
are generated from either {\tt{MC-Glauber}} or {\tt{MC-KLN}} model~\cite{Drescher:2006ca,Hirano:2009ah}.
Although, with these two initial conditions, the hybrid model simulations
successfully describe the elliptic flow of all charged and identified hadrons at RHIC and the
LHC~\cite{Song:2010mg,Song:2013qma,Zhu:2015dfa,Zhu:2015eea},
they fail to simultaneously fit all the flow harmonics $v_n$ (n=2-5) with one constant $\eta/s$ and other fine tuned parameters~\cite{Qiu:2012uy,Shen2015}. Recently, Ref.~\cite{Bhalerao:2015iya} has shown that,
with the fluctuating initial conditions generated from {\tt{AMPT}},
event-by-event viscous hydrodynamics ({\tt VISH2+1}) could nicely describe the integrated and differential flow harmonics $v_2$, $v_3$, $v_4$ and $v_5$
of all charge hadrons in 2.76 A TeV  Pb+Pb collisions. Following that paper~\cite{Bhalerao:2015iya},
we implement a string melting {\tt{AMPT}} model (version 2.21)~\footnote{Following~\cite{Xu:2011fi}, the related parameters in AMPT are set as: Lund string fragmentation parameters $a=0.5$ and $b=0.9$, strong coupling constant $\alpha=0.33$, and the screening mass $\mu=3.2 \mathrm{fm}^{-1}$.} to generate the needed initial conditions. We assume the system reaches thermalization at the hydrodynamic starting time $\tau_0$.
The corresponding initial energy density profiles in the transverses plane are constructed from the deposed
energies of produced partons from {\tt AMPT} (within space-time rapidity $|\eta_{s}|<1$) with a Gaussian smearing~\cite{Pang:2012he,Bhalerao:2015iya}:
\begin{equation}
	\epsilon(x,y) = K\sum_{i}\frac{E_{i}^{*}}{2\pi\sigma^{2}\tau_{0}\Delta\eta_{s}}\exp\left.(-\frac{(x-x_{i})^{2}+(y-y_{i})^{2}}{2\sigma^2}\right.), \label{eq:epsilon}
\end{equation}
where $\sigma$ is the Gaussian smearing factor, $E_{i}^{*}$ is the Lorentz invariant energy of the produced partons
and $K$ is an additional normalization factor.
Here, we neglect the initial flow~\footnote{Using (3+1)-d ideal hydrodynamics, Pang and his collaborators
have demonstrated that initial flow from AMPT does not significantly influence the soft hadron data like the $p_T$-spectra and the elliptic flow~\cite{Pang:2012he}.}
because one can not precisely construct both initial flow and initial energy density
for (2+1)-d hydrodynamics with an exact boost invariance from the (3+1)-d {\tt{AMPT}} simulations
without such boost invariance. We thus truncate the total produced partons of {\tt AMPT} 
within $|\eta_{s}|<1$ ($\Delta \eta_{s}=2$) and then
construct the needed boost invariant initial energy density profiles in the transverse plane
according to Eq.~(1).

In the following calculations, the hydrodynamic starting time $\tau_0$ is set to $0.4$ fm/$c$ with
the normalization factor $K$ slightly tuned around 1 to fit the multiplicity
and $p_T$ spectra of all charged hadrons in the most central Pb+Pb collisions.
We also find that, with the Gaussian smearing parameter and the specific shear viscosity
respectively tuned to $\sigma=0.6$ fm
and  $\eta/s = 0.08$, {\tt{iEBE-VISHNU}} with {\tt{AMPT}} initial conditions could
simultaneously fit the differential flow harmonics $v_2(p_T)$, $v_3(p_T)$ and $v_4(p_T)$ of all charged hadrons
at various centralities, especially from semi-central to semi-peripheral Pb+Pb collisions. For simplicity,
we neglect the bulk viscosity, net baryon density, and heat conductivity as most of the hydrodynamics and hybrid model calculations
at the LHC energies~\cite{Ryu:2015vwa,Niemi:2015qia,Schenke:2010rr,Gale:2012rq,Song:2013qma,Zhu:2015dfa}. Following Ref.~\cite{Zhou:2015iba}, the differential flow harmonics of identified hadrons are calculated with the Q-cumulant methods~\cite{Bilandzic:2010jr,Bilandzic:2013kga,Zhou:2015iba} with a pseudo-rapidity gap
$|\Delta\eta|>0$.

\begin{figure*}[!htb]
	\begin{centering}
 	\includegraphics[scale=0.95]{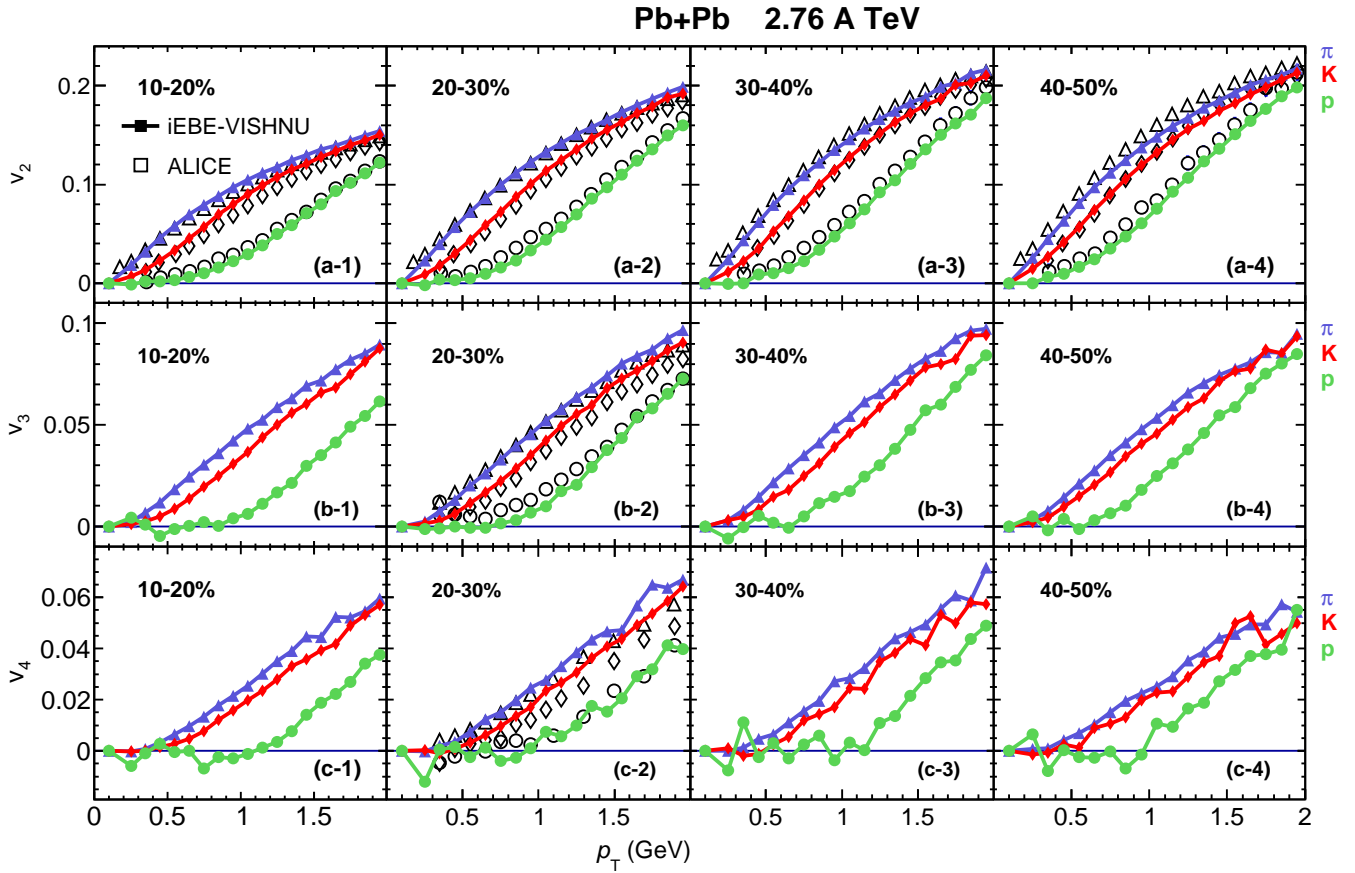}
	\end{centering}
    \vspace{-7mm}
	\caption{(Color online) $v_{n}(p_T)$ ($n=2,3,4$) of pions, kaons and protons in 2.76 A TeV Pb+Pb collisions at $10-20\%$, $20-30\%$, $30-40\%$ and $40-50\%$ centralities. The experimental data are from ALICE~\cite{Abelev:2014pua,ALICE-QM2015}, and the theoretical curves are calculated from {\tt{iEBE-VISHNU}} with {\tt{AMPT}} initial conditions.
	\label{fig:flowharmonics} }
\end{figure*}

\begin{figure}[htb]
	\begin{centering}
 	\includegraphics[scale=0.35]{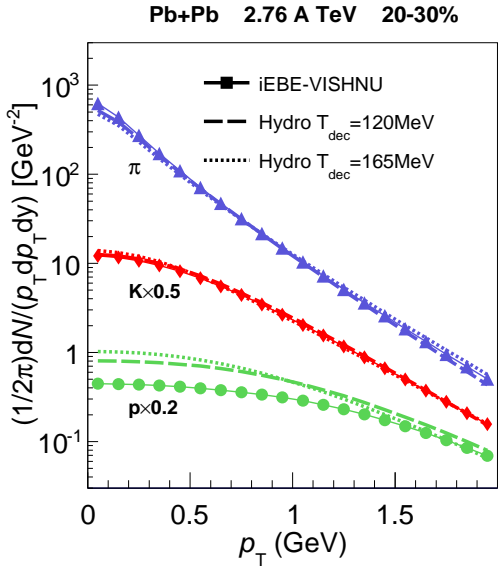}
	\end{centering}
    \vspace{-7mm}
	\caption{(Color online) $p_{T}$-spectra of pions, kaons and protons in 2.76 A TeV Pb+Pb
collisions at $20-30\%$ centrality, calculated from {\tt iEBE-VISHNU} and from
{\tt VISH2+1} with $T_{dec}= 165 \ \mathrm{MeV}$ and $T_{dec}= 120 \ \mathrm{MeV}$.
	\label{fig:flowharmonics} }
\end{figure}

\begin{figure*}
	\begin{centering}
 	\includegraphics[scale=0.95]{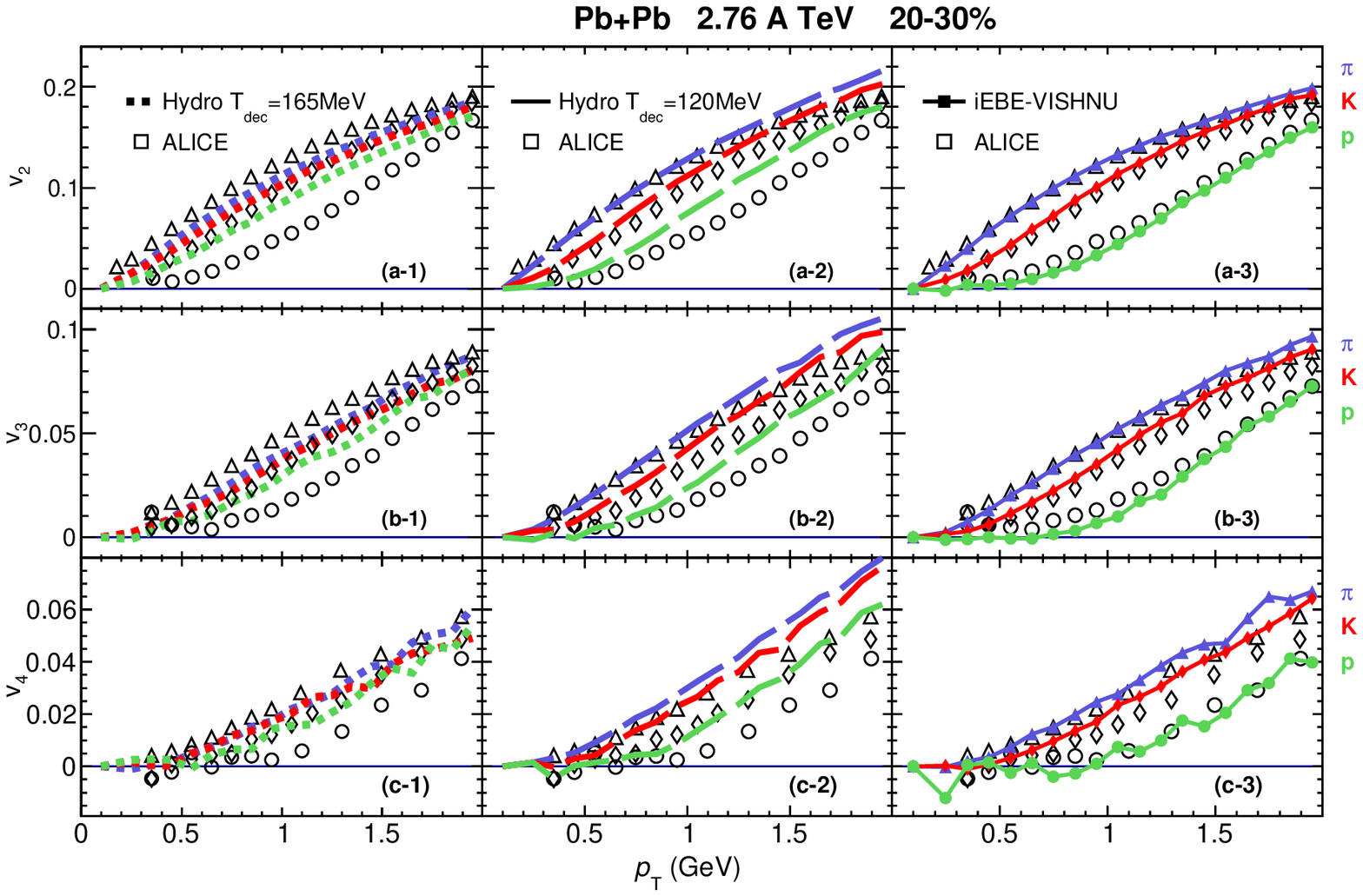}
	\end{centering}
    \vspace{-7mm}
	\caption{(Color online) $v_{n}(p_T)$ ($n=2,3,4$) of pions, kaons and
    protons in 2.76 A TeV Pb+Pb collisions at $20-30\%$ centrality, calculated from {\tt VISH2+1}
    with $T_{dec}= 165 \ \mathrm{MeV}$ (left panels) and $T_{dec}= 120 \ \mathrm{MeV}$
    (middle panels) and from {\tt{iEBE-VISHNU}} (right panels).
	\label{fig:comparison} }
\end{figure*}

\section{Results and discussions}
\label{results}

Fig.~1 shows the $p_{T}$-spectra of pions, kaons and protons in 2.76 A TeV Pb-Pb collisions at $10-20\%$, $20-30\%$,
$30-40\%$ and $40-50\%$ centralities. After tuning the starting time $\tau_0$ to 0.4 fm/c, {\tt{iEBE-VISHNU}} with {\tt{AMPT}} initial conditions
nicely fits the ALICE data, especially for the slope of the $p_{T}$-spectra of various hadron species, which indicates that {\tt{iEBE-VISHNU}}
generates a proper amount of radial flow during its QGP and hadronic expansion. We have also noticed that the $p_{T}$-spectra of pions in
our model calculations are slightly below the experiment data at 30-40\% and 40-50\% centralities because {\tt{iEBE-VISHNU}}
with {\tt{AMPT}} initial conditions under-predicts the multiplicity of all charged hadrons
in semi-peripheral and peripheral collisions~\footnote{Instead of using the empirical formula of {\tt AMPT} to cut the centrality as used in~\cite{Bhalerao:2015iya,Xu:2011fi} (which gives better descriptions of the centrality-dependent soft hadron data with the 
typical {\tt AMPT} parameter sets), we define the centrality bins
from the distributions of initial entropy, which leads to the slight under-predictions of multiplicity, spectra of all charged hadrons for semi-peripheral and peripheral collisions. Since the related effects on the differential flow are pretty small, we do not further fine-tuning the typical parameters inside {\tt AMPT}~\cite{Xu:2011fi} to achieve better descriptions for the centrality dependent multiplicity and spectra, but leave it to the future study.}.

Fig.~2 shows the differential flow harmonics $v_2(p_T)$, $v_3(p_T)$ and $v_4(p_T)$ of pions, kaons and protons
in 2.76 A TeV Pb-Pb collisions. The ALICE $v_2(p_T)$ data at 10-20\%, 20-30\% and 30-40\% and 40-50\% centralities
are measured with the scalar product method with a pseudorapidity cut
$|\Delta\eta|>0.9$~\cite{Abelev:2014pua}. The $v_3(p_T)$ and $v_4(p_T)$ data at 20-30\% centrality bin are also from ALICE,
but measured with the Q-cumulant method ($|\Delta\eta|>0$). The theoretical curves are
calculated from {\tt{iEBE-VISHNU}} with {\tt{AMPT}} initial conditions and the parameter sets described in Sec.~II. Since this paper aims
to study and predict the higher order flow harmonics of identified hadrons for the ALICE experiment, we calculate 
$v_{n}(p_T)$ from {\tt iEBE-VISHNU} simulations with the Q-cumulant method with a zero pseudorapidity
cut ($|\Delta\eta|>0$) following Ref.~\cite{ALICE-QM2015}.

For 20-30\% centrality, these ALICE data in Fig.~2 show that $v_3(p_T)$ and $v_4(p_T)$
present similar mass orderings among pions, kaons and protons as the $v_2$ mass 
ordering once observed in different colliding systems at RHIC and the LHC~\footnote{The ALICE
collaboration has also measured $v_n(p_T)$ of identified hadrons in ultra-central
collisions at 0-1\% centrality, which show similar mass ordering patterns as 20-30\% centrality.}.
With $\eta/s$ and $\sigma$ fine tuned to fit $v_n(p_T)$ of
all charged hadrons, {\tt{iEBE-VISHNU}} with {\tt{AMPT}} initial conditions  nicely describes
$v_2(p_T)$, $v_3(p_T)$ and $v_4(p_T)$ of identified hadrons at 20-30\% centrality,
especially for the mass ordering among pions, kaons and protons. In Fig.~2, we also
predict $v_3(p_T)$ and $v_4(p_T)$ of pions, kaons and protons at
$10-20\%$, $30-40\%$, and $40-50\%$ centrality bins, which show similar mass orderings patterns
as calculated/measured at 20-30\% centrality~\footnote{The flow harmonics in
ultra-central collisions are very computational expansive for {\tt iEBE-VISHNU} simulations.
Previous studies have also shown that integrated $v_n$ (n=2, 3, 4, 5, 6) of all charged hadrons
can not be simultaneously fitted without further considering the effects from bulk viscosity,
nucleon-nucleon correlations in the initial state, etc.~\cite{Denicol:2014ywa}. Therefore, we will not further
explore $v_n(p_T)$ of identified hadron at 0-1\% centrality as already measured by ALICE
in our current investigations.}.

To explore in depth the development of flow anisotropy and its distribution among various hadron species during
the hadronic evolution, we further calculate the $p_T$ spectra and differential flow harmonics of
identified hadrons at 20-30\% centrality with three comparison runs, using the same AMPT initial conditions
and parameter sets: (1) / (2) viscous hydrodynamics simulations +
resonance decays, with the decoupling temperature set to
 $T_{dec}=165\ \mathrm{MeV}$ / $T_{dec}=120\ \mathrm{MeV}$, (3) {\tt{iEBE-VISHNU}} simulations that combines
viscous hydrodynamics with the {\tt UrQMD} hadronic afterburner, which are the same simulations
as presented in Fig.~1 and Fig.~2 for 20-30\% centrality. Please note that $T_{dec}=165 \ \mathrm{MeV}$ is also the switching temperature between the
macroscopic and  microscopic approaches in the {\tt{iEBE-VISHNU}}
simulations. With $T_{dec}=165 \ \mathrm{MeV}$, we therefore eliminate
the hadronic evolution in the hydrodynamic simulations and concentrate on the flow developed in the
QGP phase. For the hydrodynamic simulations with $T_{dec}=120 \ \mathrm{MeV}$, the hadronic matter below $T_c$ expands
like a fluid, which keeps the local thermal equilibrium till the kinetic freeze-out. Meanwhile, the specially
constructed equation of state s95p-PCE maintains the partially chemical equilibrium of the system during the
hadronic evolution. In contrast, {\tt iEBE-VISHNU}
below $T_{sw}$ simulates the microscopic expansion of a hadron resonance gas with both scatterings and decays,
which quickly evolves out of equilibrium and finally becomes free-streaming hadrons in the very late
stage of evolution.

Fig.~3 shows that, for pure hydrodynamic simulations with EoS s95p-PCE, the calculated spectra of pions and kaons are not
sensitive to the decoupling temperature $T_{dec}$, both of which are almost overlap with the counterparts
from {\tt iEBE-VISHNU}. The spectra of protons from hydrodynamic simulations become flatter as $T_{dec}$
decreased from 165 MeV to 120 MeV, indicating the development of additional radial flow during the hadronic fluid expansion.
The {\tt UrQMD} hadronic evolution below $T_{sw}$ also build up additional radial flow. Meanwhile, the baryon anti-baryon
annihilations tend to eat-up the (anti-)protons at lower-$p_T$. With these two factors, {\tt iEBE-VISHNU} obtains
a flatter and lower $p_T$ spectra of protons when compared with the ones from pure hydrodynamic simulations.

Fig.~4 presents the differential flow harmomics $v_2(p_T)$, $v_3(p_T)$ and $v_4(p_T)$
of identified hadrons in 20-30\% Pb+Pb collisions, calculated from viscous hydrodynamics with $T_{dec}=165 \ \mathrm{MeV}$
and $T_{dec}=120 \ \mathrm{MeV}$, and from {\tt iEBE-VISHNU}. In each panel, the theoretical calculations 
are respectively compared with the ALICE data.  The {\tt iEBE-VISHNU}
results in the  right panels are exactly the same as the corresponding ones shown in Fig.2,
which nicely fit the ALICE data of
pions, kaons and protons at 20-30\% centrality. In contrast, pure hydrodynamics simulations
without hadronic evolution ($T_{dec}=165 \ \mathrm{MeV}$, left panels) under-predict the pion's
$v_2$, $v_3$ $v_4$ due to the insufficient flow anisotropy development in the
QGP phase. Meanwhile, the smaller radial flow there (indicated by Fig.~3) also
leads to smaller mass-splittings between pions and protons for different flow harmonics.
With $T_{dec}$ decreased to $120 \ \mathrm{MeV}$, the hydrodynamics simulations (middle panels)
roughly fit the ALICE data for pions and show  enhanced
the mass-splittings between pions and protons for $v_2$, $v_3$ and $v_4$ due to
the additional collective flow developed in the hadronic stage. However,
the mass-splittings are still under-predicted when compared with the ALICE data
since $v_2(p_T)$, $v_3(p_T)$ and $v_4(p_T)$ of protons are still well above the experimental
data. Such situation is similar to the case of elliptic flow as found in Ref.~\cite{Shen:2011eg}, which showed
that, although pure hydrodynamics could nicely fit many soft hadron data of
all charged hadrons and pions, the elliptic flow
of protons below 2 GeV are still over predicted in central and semi-central
Pb+Pb collisions. A later calculations from {\tt VISHNU} hybrid model~\cite{Song:2013qma} pointed out
that the hadronic scatterings in the {\tt UrQMD} evolution rebalance the generation of
radial and elliptic flow, leading to improved descriptions of the proton $v_2$ as well as the mass-splitting of $v_2$ between pions and protons.
The comparison runs in Fig.~4 here demonstrate that microscopic hadronic scatterings in {\tt iEBE-VISHNU} are also important for the nice fit of
$v_3(p_T)$ and $v_4(p_T)$ of protons, which leads to an improved description
of the mass-splitting between pions and protons for these higher order flow harmonics.

\section{Summary}
\label{Summary and remarkable conclusions}
Using the event-by-event {\tt VISHNU} hybrid model {\tt iEBE-VISHNU} that couples 2+1-d viscous hydrodynamics to the
hadron cascade model {\tt UrQMD}, together with the fluctuating {\tt AMPT} initial conditions, we studied
the $p_{T}$-spectra, differential flow harmonics $v_2(p_T)$, $v_3(p_T)$ and $v_4(p_T)$ of identified hadrons in
2.76 A TeV Pb+Pb collisions.  With fine tuned parameters, our model calculations
 nicely described the ALICE data at 20-30\% centrality, especially for the mass-orderings among pions, kaons and
protons. We also predicted the higher order flow harmonics of identified hadrons at other centrality bins,
which showed a similar mass-ordering pattern as measured/calculated
at 20-30\% centrality.

We explore the development of $v_n$ mass orderings/splittings during the hadronic evolution through
the comparison runs from {\tt iEBE-VISHNU} simulations and from pure hydrodynamic simulations decoupled at
different temperatures. We found that, in pure hydrodynamics, the additional radial flow developed in
the fluid-like hadronic phase could enhance mass splitting between pions and
protons for all $v_n$ (n=2, 3, 4). However, the protons data are still over-predicted there.
In contrast, the microscopic hadronic scatterings in {\tt iEBE-VISHNU} rebalance the
generation of radial and anisotropy flow, leading to nice descriptions of the
$v_2$, $v_3$ and $v_4$ data for pions, kaons and protons.

At last, we would like to remind the interested readers that the past research on elliptic flow~\cite{Song:2013qma} has showed,
although {\tt VISHNU} hybrid model improves the description of elliptic flow for pions,
kaons and protons from most central to semi-central collisions, it still slightly under-predicts the proton $v_2$ 
in semi-peripheral collisions. Further investigations on strange and multi-strange hadrons~\cite{Zhu:2015dfa,Zhu:2015eea,Heinz:2015arc}
also showed that {\tt VISHNU} simulations can not correctly
reproduce all the $v_2$ mass orderings among various stable hadrons, especially for the mass ordering
between protons and Lambdas. Currently, the ALICE
collaboration only release the measurement of higher order flow harmonics of pions, kaons and protons
in ultra-central (0-1\%) and semi-central (20-30\%) Pb+Pb collisions at $\sqrt{s_{NN}} = $ 2.76 A TeV.
In fact, extending the current measurements to various centrality bins and collision energies, and even to rarely produced
strange and muti-strange hadrons will further test our model calculations, broadening our knowledge on the
hot and evolving hadronic matter after the QCD phase transition.

\section*{Acknowledgments}
We thank R.~S.~Bhalerao, L.~Pang and X.~Zhu for valuable discussions. This work is supported by the NSFC and the MOST under grant
Nos.11435001 and 2015CB856900. H.~X. is partially supported by China Postdoctoral Science Foundation with
grant No.~2015M580908. We gratefully acknowledge the extensive computing resources provided to us
by Supercomputing Center of Chinese Academy of Science (SCCAS) and Tianhe-1A from the 
National Supercomputing Center in Tianjin, China.

\end{document}